\documentclass[12pt,preprint]{aastex}

\begin{document}


\title{A Classic Type 2 QSO}


\author {Colin Norman\altaffilmark{1,2}, Guenther
Hasinger\altaffilmark{3,4}, Riccardo Giacconi\altaffilmark{1,5}, Roberto
Gilli\altaffilmark{1,6}, Lisa Kewley\altaffilmark{1,7}, Mario
Nonino\altaffilmark{8}, Piero Rosati\altaffilmark{1,9}, Gyula
Szokoly\altaffilmark{3}, Paolo Tozzi\altaffilmark{8}, Junxian
Wang\altaffilmark{1}, Wei Zheng\altaffilmark{1}, Andrew
Zirm\altaffilmark{1}, Jacqueline Bergeron\altaffilmark{9}, Roberto
Gilmozzi\altaffilmark{9}, Norman Grogin\altaffilmark{2}, Anton
Koekemoer\altaffilmark{2}, and Ethan Schreier\altaffilmark{2}}

\altaffiltext{1}{The Johns Hopkins University, Homewood Campus, Baltimore, MD 21218}

\altaffiltext{2}{Space Telescope Science Institute, 3700 San Martin Drive, Baltimore, MD 21218}

\altaffiltext{3}{Astrophysikalisches Institute Potsdam, An der Sternwarte 16, Potsdam, D-14482, Germany}

\altaffiltext{4}{Max-Planck-Institut f\"ur Extraterrestrische Physik, Giessenbachstrasse, Garching, D-85740, Germany}

\altaffiltext{5}{Associated Universities, Inc. 1400 16th Street, NW, Suite 730, Washington, DC 20036}

\altaffiltext{6}{Osservatorio Astrofisico di Arcetri, Largo E. Fermi
5, 50125 Firenze, Italy}

\altaffiltext{7}{Harvard-Smithsonian Center for Astrophysics, 60 Garden Street, Cambridge, MA 02138}

\altaffiltext{8}{Osservatorio Astronomico, Via G. Tiepolo 11, 34131 Trieste, Italy}

\altaffiltext{9}{European Southern Observatory, Karl-Schwarzschild-Strasse 2, Garching, D-85748, Germany}

\begin{abstract}
 
In the {\it Chandra} Deep Field South 1Msec exposure we have found, at
redshift $3.700 \pm 0.005$, the most distant Type 2 AGN ever detected.
It is the source with the hardest X-ray spectrum with redshift $z>3$.
The optical spectrum has no detected continuum emission to a 3$\sigma$
detection limit of $\sim 3 \times 10^{-19}$ ergs/s/cm$^{2}$/\AA\ and
shows narrow lines of Ly$\alpha$, CIV, NV, HeII, OVI, [OIII], and
CIII].  Their FWHM line widths have a range of $ \sim 700 - 2300$ km
s$^{-1}$ with an average of approximately $\sim 1500$ km s$^{-1}$. The
emitting gas is metal rich ($Z\simeq 2.5 - 3 Z_\odot$).

In the X-ray spectrum of 130 counts in the $0.5-7$ keV band there is
evidence for intrinsic absorption with $N_H\gtrsim 10^{24}$
cm$^{-2}$. An iron K$\alpha$ line with rest frame energy and
equivalent width of $\sim6.4$ keV and $\sim 1$ keV, respectively, in
agreement with the obscuration scenario, is detected at a $2\sigma$
level. If confirmed by our forthcoming XMM observations this would be
the highest redshift detection of FeK$\alpha$. Depending on the
assumed cosmology and the X-ray transfer model, the 2-10 keV rest
frame luminosity corrected for absorption is $\sim 10^{45 \pm 0.5}$
ergs s$^{-1}$, which makes our source a classic example of
the long sought Type 2 QSOs. From standard population synthesis
models, these sources are expected to account for a relevant fraction
of the black-hole-powered QSO distribution at high redshift.

\end{abstract}

\section{Introduction}
                      
The unified model for AGNs is widely accepted. Briefly, the physics of
black hole, accretion disk, jet, and obscuring torus is convolved with
the geometry of the viewing angle and can explain most of the apparent
disparate properties of active galaxies \citep{antonucci1993}. The use
of the word torus here is generic for the obscuring region since there
are many variants of the geometry of the obscuring region including a
strict toroidal geometry and flaring disk models \citep{efstathiou95,
granato97}. Type 1 objects exhibit the straight physics of AGNs with
no absorption and Type 2 objects arise when the view is obscured by
the torus. A crucial component that has long been sought are heavily
obscured powerful quasars called QSO 2s.  They have been predicted to
have narrow permitted lines, powerful hard X-ray emission and a high
equivalent width Fe K$\alpha$ line \citep{ghisellini1994}.

In the absence of a good standard case doubt has been expressed, at
times, about the whole QSO 2 phenomenon \citep{halpern1999}. The
successful finding of an optically obscured central AGN in the
starburst NGC 6240 \citep{vignati} made X-rays the obvious
wavelength to uncover the Type 2 QSO phenomenon.  There have been few
previous studies of candidate Type 2 objects in the X-ray band. One
has been observed by ROSAT and ASCA (Almaini et al. 1995;
Georgantopoulos et al. 1999).  Two are underluminous objects from ASCA
\citep{ohta1996} and {\it Chandra} \citep{fabian2000}. Also,
the hyper-luminous infrared galaxy IRAS 09104 +4109,
observed with BeppoSAX \citep{franceschini2000} and recently with {\it
Chandra} \citet{iwasawa2001}, appears to be an example of a Type 2 QSO
at moderate redshift, $ z =0.442$. It is a cD galaxy in a rich cluster
and is radio loud with a radio jet.  A recent candidate at high
redshift ($z=3.288$) has been found in the deep Chandra exposure of
the Lynx field (Stern et al. 2001).  

In parallel with the X-ray work, imaging and spectropolarimetric
studies on warm ultraluminous, infrared galaxies (ULIRGs) indicate
that all of them might contain buried QSOs \citep{tran00}.  In
addition, high redshift radio galaxies (HZRGs) harbour obscured QSOs,
and intensive studies are underway to understand the physics of the
central regions of these objects \citep{Vernet01}.

From our multiwaveband studies of the 1Msec exposure of the {\it
Chandra} Deep Field South (CDF-S) (Giacconi et al. 2001a; Tozzi et
al. 2001; Rosati et al. 2001; Giacconi et al. 2001b) we now discuss a
classic X-ray selected Type 2 QSO: CXOCDFS J033229.9 -275106
(hereafter CDF-S 202). We note in passing that, in contrast to the
above mentioned radio studies, CDF--S 202 is radio quiet.
 
The Paper is organized as follows. In Section II we describe the
multiwavelength observations we use. In Section III we
present and analyze the optical spectrum. In Section IV we discuss the
X-ray spectrum. In Section V we study the object further and present
some of the wider implications of this work. We conclude and consider
future work in the final section.

We assume a cosmology in this paper with the parameters $H_0=60$ km
s$^{-1}$Mpc$^{-1}$, $\Omega_m=0.3$ and $\Omega_{\Lambda} = 0.7$.

\section{Imaging Observations} \label{observations}

We have combined 11 individual {\it Chandra} pointings into a mosaic
with a maximum of 942 ksec of exposure time (hereafter the 1 Msec
exposure).  The limiting fluxes achieved are $5.5 \times 10^{-17}$ erg
cm$^{-2}$ s$^{-1}$ in the 0.5-2 keV band and $ 4.5 \times 10^{-16}$
erg cm$^{-2}$ s$^{-1}$ in the 2-10 keV band (Rosati et al. 2001).  The
details of the X-ray observations are given in Giacconi et al. (2001b,
submitted).  Among the sources with $z>3$, CDF--S 202 is the one
with the hardest X-ray spectrum. In the X-ray image from $0.5-7$ keV
the source is $\sim3\arcmin$ from the center, where the point response
function has a full--width--half--maximum (FWHM) of 1$\arcsec$. The X--ray
intensity profile is completely consistent with that of a point--source
at this resolution.  There are no other X-ray sources detected within
30$\arcsec$.

\begin{figure}
\plotone{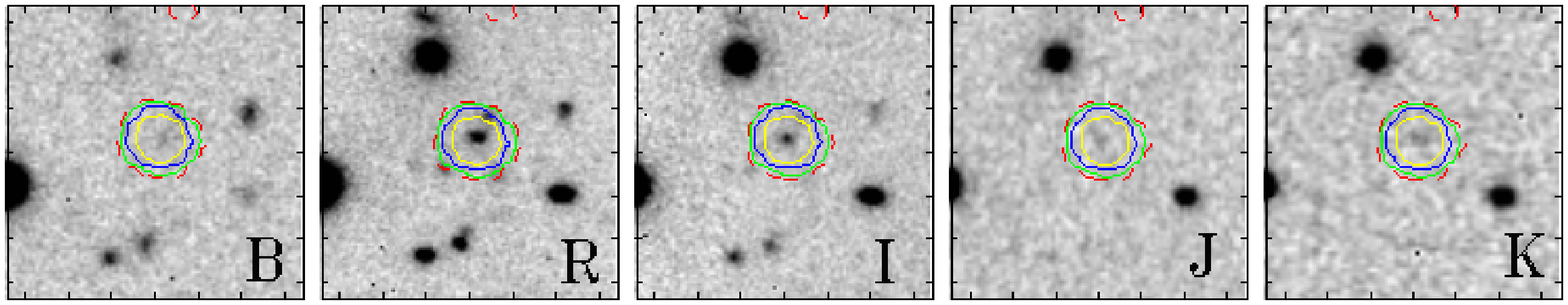}
\caption{Optical and near IR images with the X-ray brightness contours
overlaid. Optical and near IR images are $20\arcsec$ across; the
X--ray contours correspond to 3,5,10 and 20$\sigma$ levels above the
background for the 2--7 keV image  
\label{cut}}
\end{figure}

The primary optical ID was made in deep $R$ and $I$ band images
($R$(Vega)$\leq26$) obtained using the Focal Reducer Spectrograph
(FORS1) on the European Southern Observatory Very Large Telescope 8.2m
facility (ESO/VLT--ANTU). All magnitudes quoted in this paper are Vega
magnitudes. CDF-S 202 has an $R$ magnitude of 23.53 and an $I$
magnitude of 22.65.  We have additional photometric coverage of this
object in the $U$, $V$, $B$, $J$ and $Ks$ bands.  The B band used here
is from ESO Wide--Field Imager data taken on the 2.2m ESO telescope,
which reaches a limiting magnitude of 26.  The near--infrared data
comes from the ESO Imaging Survey (EIS) and reaches 23.6 in $J$ and
21.8 in $Ks$.  Of these five bands our candidate QSO 2 is detected
significantly in $V$ and $Ks$, at 25.27 and 20.99 respectively.  The
$B$--band source is very faint, essentially at the limiting
magnitude. The images in the $B$, $R$, $I$, $J$ and $Ks$ bands are
given in Fig.~\ref{cut}.

\section{Optical Spectroscopy} \label{optical}

\subsection{Observations and Data Reduction}

The optical spectrum was obtained with the multislit mode on the Focal
Reducer Spectrograph (FORS1) on the European Southern Observatory Very
Large Telescope 8.2m unit facility (ESO/VLT--ANTU) on 11--25--2000.
The FORS instrument is described in detail in \citet{Mitsch94}.  The
spectroscopy was obtained using the 150I+17 grism and no order
separation filter, which allows maximum wavelength coverage but
introduces second order contamination.  This grism gives a spectral
resolution of 5.5\AA$/$pixel.  The detector slit width was 1.2
arcseconds, (around 30\AA\  dispersed), chosen to maximize the
signal-to-noise ratio of the resulting spectra (i.e. including most of
the light from the object, but minimizing the sky background).  The
final spectrum is composed of 7 exposures with a total integration
time of $\sim$3 hours.  

Data reduction was carried out using the IRAF package.  This involved
bias subtraction, flat fielding, background subtraction, wavelength
and flux calibration.  The background was subtracted by fitting a
second order polynomial to each column.  The spectrum was extracted
using an aperture width of 14 pixels, which was sufficent to include
the total emission from the object.

Wavelength calibration was derived using the spectra from four arc
lamps (He, HgCd and two separate Ar lamps) taken on the same night.  Line centers and shapes
can only be determined to a few \AA\  accuracy due to the finite slit
width. In addition, the exact location of the object on the slit can
introduce an additional few \AA\ {\em systematic} shift in the
spectrum. These limit the redshift to an accuracy of $\pm0.005$.  It
is also important to note that the finite slit size and seeing
introduces an artificial broadening of the lines. As the seeing is
potentially wavelength dependent, this effect can be also wavelength
dependent. We estimate this effect to be a few times 10\AA. Therefore,
the line widths measured (see Table 1) should be considered upper
limits.

The spectra were flux calibrated  using very wide (5 arcsec) slits
using both a red and a blue photometric standard star for comparison.
As the seeing is potentially wavelength dependent, applying this flux
calibration can introduce a wavelength dependent slit-loss (in
addition to  the varying slit loss introduced by the atmospheric
refraction). Considering the relatively good seeing (0.5" FWHM at the
beginning of the observation) and
low airmass, this color effect  is not very strong.  
Second order contamination was removed {\em from the
specphot standard observations} using two standards with very different
spectral shape. This made it possible to calibrate the true throughput
of the system. Due to the very faint continuum of CDF-S 202, the second
order contamination was not removed from its spectra. This implies that
the continuum shape would be incorrect (if we could detect it), but the
emission line features are correctly calibrated.
Comparing the
spectra obtained using each photometric standard, we estimate that our
flux calibration is accurate to 5\%.  At present, we use known CTIO
extinction curves until ESO extinction curves for Paranal become
available to us.

\subsection{Emission-Line Analysis}

QSOs can be classified into Type 2 and Type 1 objects in a similar
fashion to Seyfert galaxies, which are generally classified based on
their relative emission--line widths using the scheme proposed by
\citet{Weedman70,Weedman73} and \citet{Khachikian71,Khachikian74}.  In
this scheme, Seyferts which show broad \ion{H}{1}, \ion{He}{1} and
\ion{He}{2}\ emission lines FWHM $\sim 3-5 \times 10^{3}$ km s$^{-1}$)
are known as Seyfert 1 galaxies.  These galaxies also display
strong blue continua, and frequently complexes of broad \ion{Fe}{2}
emission are also seen.  Superimposed on this spectrum are `narrow'
forbidden lines, which are thought to be formed in a much larger
extended narrow line region (NLR) around the nucleus.

The forbidden lines in Seyfert 1 galaxies such as [\ion{O}{3}],
[\ion{N}{2}], and [\ion{S}{2}] typically have FWHM of $\sim 5 \times
10^{2}$ km s$^{-1}$.  Galaxies with permitted and forbidden lines with
approximately the same FWHM (typically $\sim 5 \times 10^{2}$ km
s$^{-1}$) are called Seyfert 2 galaxies.  The broad lines are
absent in such objects and they display a flat, featureless continuum
\citep[eg.,][]{Kinney93,heckman1995}.  Similarly, broad QSO emission
lines typically have widths of 3000-5000 km s$^{-1}$, while narrow QSO
emission lines are a few hundreds of km s$^{-1}$
\citep[eg.,][]{Peterson97,Forster01}.

\begin{figure}
\plotone{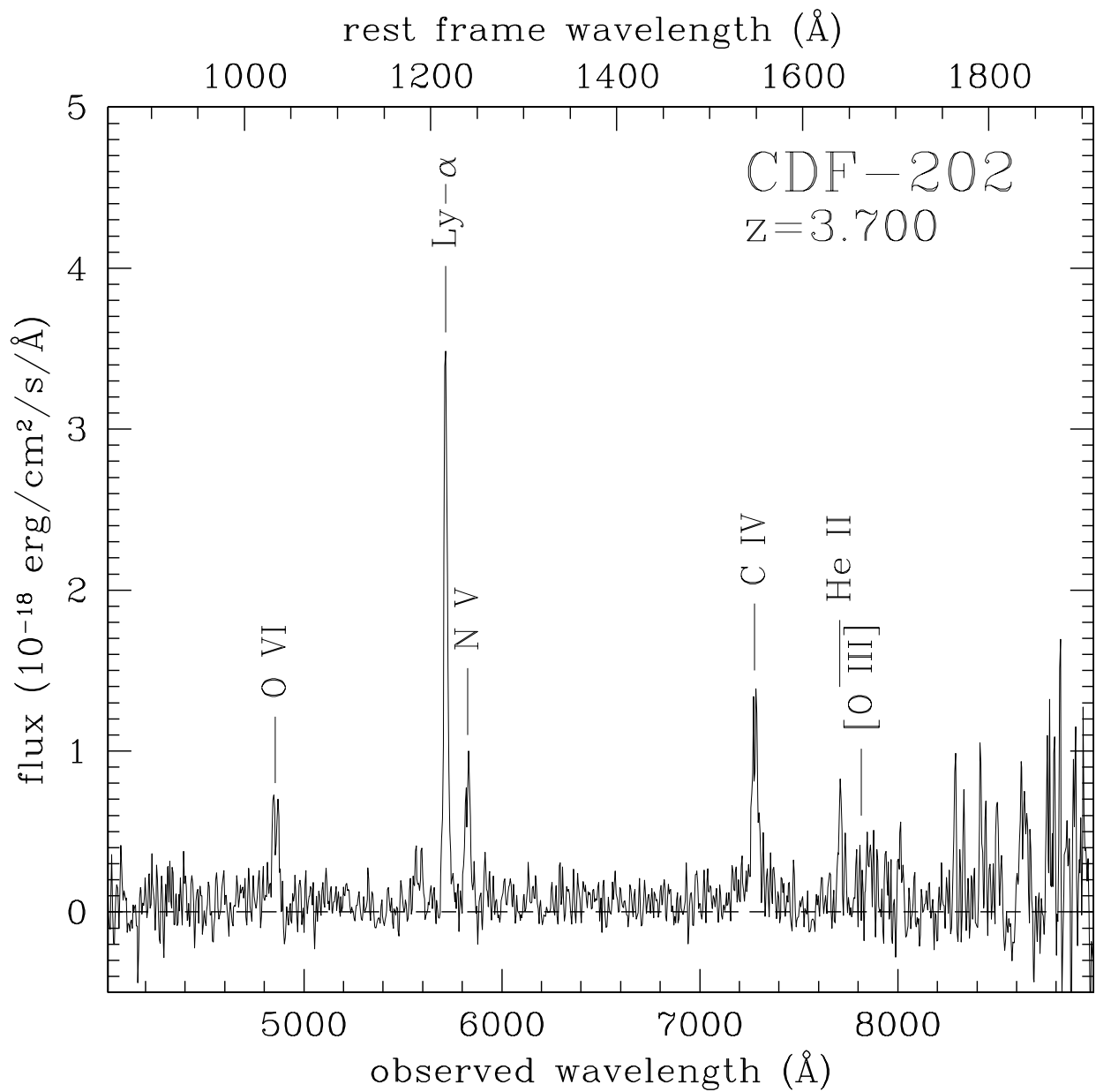}
\caption{Low resolution optical spectrum of CDF-S 202 obtained with
VLT UT1 FORS1. The emission features that the redshift determination
is based on are marked. Wavelength calibration inaccuracies limit the
accuracy of redshift determination to $\pm$0.005. Flux calibration of
emission lines is good to 5\% in the range displayed. Data were
not corrected for slit loss, which we estimate to be around
30\% and nearly achromatic.
\label{OptSpec}}
\end{figure}

To determine the emission line fluxes and FWHM for CDF--S 202, the
optical spectral line profiles were modeled by one Gaussian per line
using the {\it ngaussfits} task in IRAF.  This routine uses least
squares fitting implemented by a downhill simplex minimization
algorithm.  The resulting emission line fluxes and FWHM found for the
emission lines in CDF--S 202 are given in Table~1. The final spectrum
is shown in Figure~\ref{OptSpec}.  The permitted emission lines,
\ion{O}{4}, Ly$\alpha$, \ion{N}{5}, \ion{C}{4}, and \ion{He}{2} have
S/N ratios $> 3\sigma$ and are well identified, giving a redshift of
$z=3.700 \pm 0.005$.  The continuum is not detected to a 3$\sigma$
limit of $\sim 3 \times 10^{-19}$ ergs/s/cm$^{2}$/\AA.  This is
consistent with the flat, featureless continua seen in many Type 2
objects such as NGC 3393 \citep[eg.][]{Kinney93,heckman1995}.

The emission line widths of CDF--S 202 are given in Table 1. They have
a rest--frame FWHM of between $\sim 700 - 2300$ km s$^{-1}$.  Note
that these FWHM are actually upper limits as discussed in the previous
section, and that the ``true'' FWHM may be even less than those quoted
here.  The FWHM of Ly$\alpha$ has an upper limit of 1100 km s$^{-1}$.
Assuming that our FWHM upper limits are close to  the true FWHM of the
emission lines (within 100-300 km s$^{-1}$), then our Ly$\alpha$ FWHM
is similar to  the FWHM of $\sim 900$ km s$^{-1}$ observed in the Type
2 QSO IRAS 09104+4109 \citep{Kleinmann88}.  Such FWHM are broader than
those seen in typical Seyfert 2s ($\sim 5 \times 10^{2}$ km s$^{-1}$),
but narrower than those observed in Seyfert 1s and broad-line QSOs
($3000 - 5000 $km s$^{-1}$).   Our FWHM are, however,
similar to those found in the so-called ``narrow-line Seyfert 1''
galaxies which have FWHM of 1000-2000 km s$^{-1}$ \citep{Goodrich89}.
However, narrow--line Seyfert 1 galaxies are generally defined as
having Balmer lines slightly broader than the forbidden lines, an
[\ion{O}{3}]~$\lambda$5007/H$\beta$ ratio less than 3, and in most
cases, the presence of high ionization iron species.  Without spectra
at optical wavelengths in the rest-frame, and better estimates of the
FWHM, we are unable to determine whether CDF-S 202 fits these criteria.

We note that the absorption seen in CDF-S 202 of $N_H\gtrsim
10^{24}$ cm$^{-2}$ is common for Type 2 objects \citep{maio,risa}
but not for Type 1s \citep{return}.  In addition, the flat,
featureless continuum to a 3$\sigma$ flux limit of $\sim 3\times
10^{-19}$ ergs/s/cm$^{2}$ supports a Type 2 classification of CDF--S
202.

\begin{figure}
\plotone{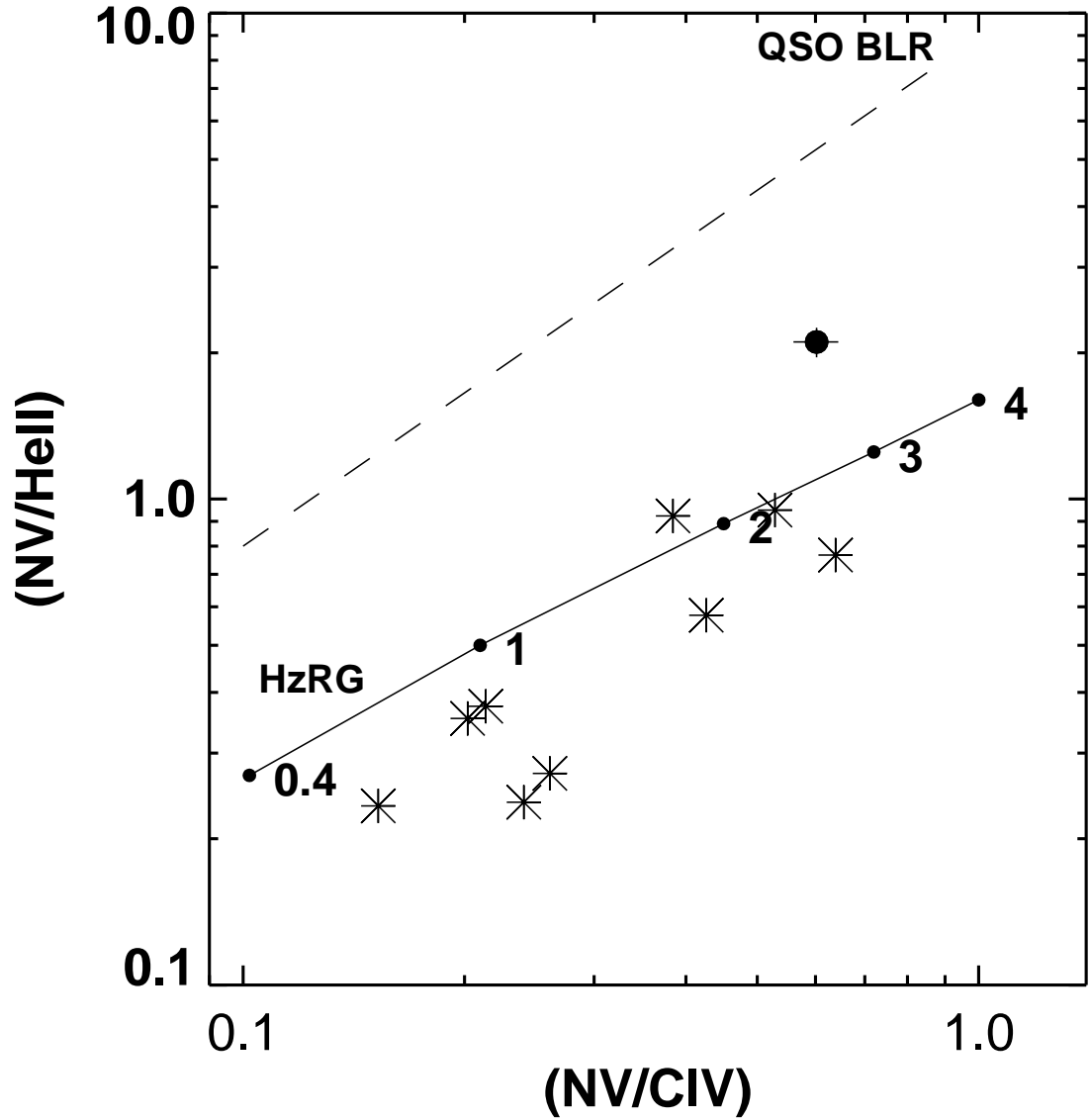}
\caption{The \ion{N}{5}/\ion{He}{2} vs. \ion{N}{5}/\ion{C}{4} diagram
as plotted by \citet{hamann93,Vernet01}.  The solid line gives the
locus of the best-fit power-law photoionization models for
metallicities ranging from $0.4-4 Z_{\odot}$.THe BLR model for QSOs
\citep{hamann93} is given by the dashed line. NLRG are shown as
asterisks and CFD-S 202 as a filled circle. CFD-S 202 appears to have
high metallicity.
\label{Linediagnostic}}
\end{figure}

Figure~\ref{Linediagnostic} shows the \ion{N}{5}/\ion{He}{2}
vs. \ion{N}{5}/\ion{C}{4} diagram with models by \citet{Vernet01} and
\citet{hamann93}.  \citet{hamann93} showed that high redshift quasars
($z>2$) define a tight correlation, indicated by the dashed line in
this figure.  This correlation is predominantly a result of
metallicity variations in which the higher redshift quasars have
higher metallicities. This redshift--metallicity effect may be due to
higher QSO or host galaxy masses at higher redshifts.  From
Figure~\ref{Linediagnostic}, we can see that the \ion{N}{5}/\ion{C}{4}
and \ion{N}{5}/\ion{He}{2} ratios for CDF--S 202 are strong, similar to
those previously seen in Type 1 QSOs and powerful radio galaxies at
similar redshifts \citep{hamann93,Vernet01}.  The position of CDF--S
202 in this figure indicates emission line ratios intermediate between
HzRGs and BLR QSOs.

The solid line shown in Figure~\ref{Linediagnostic} gives the best--fit
power-law photoionization model for high redshift radio galaxies
(HzRG) run by Vernet et al. for metallicities from $0.4-4 Z_{\odot}$.
Similarly, the \citet{hamann93} model for the broad emission-line region of
QSOs (QSO BLR) is shown in dashes. From these models, CDF-S 202
appears to be a high metallicity (Z$\sim 2.5 - 3\,\times Z_\odot$) object,
at the high end of the metallicity range found in HzRGs by Vernet et al.
although an independent metallicity diagnostic is required to verify
this.

The \ion{He}{2}/\ion{C}{4} ratio is $\sim 0.3$.  \citet{heckman1995}
note that this ratio is typically 0.1 in Seyfert 1 galaxies compared
with 0.9 for Seyfert 2 galaxies. Narrow--line radio galaxies (the
radio--loud analogs to Seyfert 2 galaxies) also show strong
\ion{He}{2}/\ion{C}{4} ratios ($\sim 0.9$) \citep{McCarthy93}, while
spectra of high-{\it z} quasars resemble the Seyfert 1 galaxies with
\ion{He}{2}/\ion{C}{4}$\sim 0.15$ \citep{Francis91}.  The
\ion{He}{2}/\ion{C}{4} ratio for CDF-S 202 is intermediate between
the Type 1 and Type 2 values. However, CDF-S 202 is more likely a
Type 2 QSO rather than an intermediate QSO, since its emission line
velocity widths are relatively narrow and heavy obscuration is
confirmed from the X--rays.

\section{X-ray Spectroscopy}  \label{x-ray}

The X--ray spectrum of CDF--S 202 and the relative background spectrum
were extracted from the X--ray image using the standard CIAO v2
software. The source spectrum was extracted in a circle with radius
$R=5 \arcsec $, the background spectrum was extracted in an annulus
between $R+2 \arcsec $ and $R+12 \arcsec $ around the source
position. We verified that changing the source extraction radius does
not affect significantly our results.  With the same software we
produced response matrix functions at the CDF--S 202 position. The
X--ray spectrum of CDF-S 202 was then analyzed using XSPEC v11.0.  The
data were rebinned to have at least 10 counts per bin (source +
background), in order to validate the use of the $\chi^2$
statistics. Hereafter errors are quoted at the 90\% confidence level
for one interesting parameter ($\Delta\chi^2=2.71$). The fit with a
simple power law absorbed by the Galactic hydrogen gives a very flat
slope ($\Gamma=0.51\pm0.27$) suggesting
heavy absorption. We then considered two possible scenarios. First, we
assumed the X--ray emission is transmitted through an absorber which
cuts off the spectrum at low energies. Usually the absorber is
identified with the putative obscuring torus expected from unified
schemes, although its geometry could be different. Second, we assumed
the X-rays are reflected. The obscuring material is then optically
thick to Compton scattering with no transmitted X-rays observed in the
{\it Chandra} band up to $7\times(1 + z) \sim 33$ keV, implying $N_{\rm
H}\ga 10^{25}$ cm$^{-2}$. In this case the X-rays are reflected by the
inner edge of the torus (or any cold cloud off the line of
sight). Generally, for large equivalent width of the Fe K$\alpha$
line, the reflection model is favoured \citep{ghisellini1994}.

\begin{figure}
\plotone{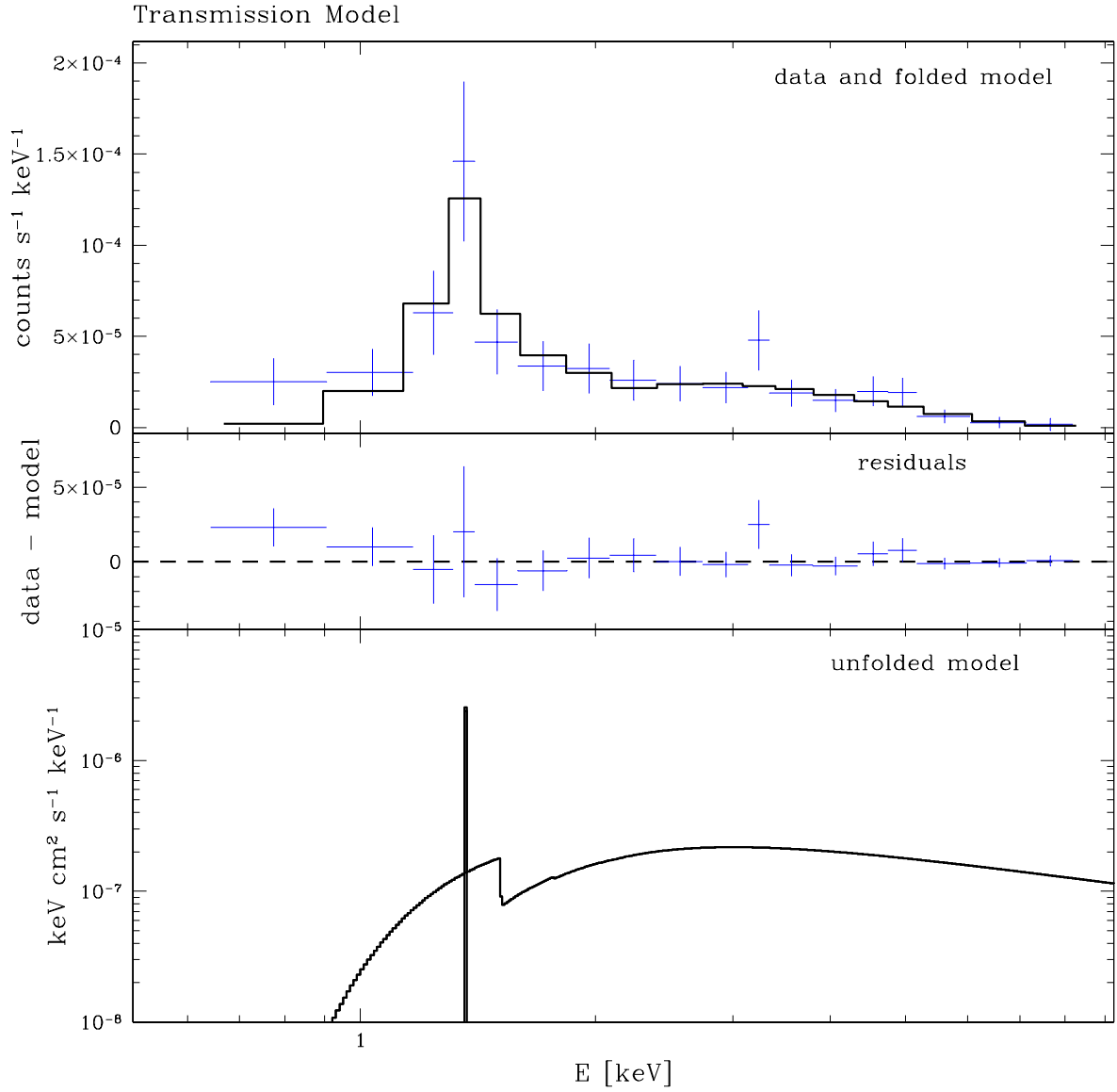}
\caption{Top panels: X--ray Spectrum of CDF--S 202 with Transmission
Fit and residuals.  Bottom panel: the unfolded best--fit model.  
\label{Xtrans}}
\end{figure}

We first considered the transmission model and fitted the spectrum
with an absorbed powerlaw. The best fit photon index ($\Gamma=1.847$)
is poorly constrained, therefore we fixed it to be $\Gamma=1.8$, a
standard value for AGN, deriving a column density of $N_{\rm H} =
6.7^{+2.4}_{-2.1} \times 10^{23}$ cm$^{-2}$. A residual excess around
1.4 keV suggested the presence of an iron K$\alpha$ line at 6.4 keV
rest frame ($6.4/(1+3.700)\simeq1.4$ keV), therefore we added a narrow
Gaussian line to the model fixing the line energy to 6.4 keV and the
line width to 0 keV. The addition of the iron line changes the
$\chi^2/dof$ value from 11.6/16 to 9.5/15; the fitted column density
slightly increases to $N_{\rm H} = 7.9^{+3.5}_{-2.3} \times 10^{23}$
cm$^{-2}$. Although the iron line detection is not significant
($<2\sigma$ level according to an $F$-test for one additional free
parameter), we will leave the line in the transmission model for
comparison with the reflection model. The line equivalent width in the
observed frame is $EW_{K\alpha}=175^{+360}_{-175}$ eV, corresponding
to $EW_{K\alpha}=823^{+1694}_{-823}$ eV in the rest frame.  The 2-10
keV rest frame luminosity is calculated to be $4.4 \times 10^{43}$ erg
s$^{-1}$.  When correcting for absorption, the intrinsic 2--10 keV
rest frame luminosity is $3.3 \times 10^{44}$ erg s$^{-1}$. In
Figure~\ref{Xtrans} we show the fit to the X--ray spectrum together
with the best fit transmission model.

\begin{figure}
\plotone{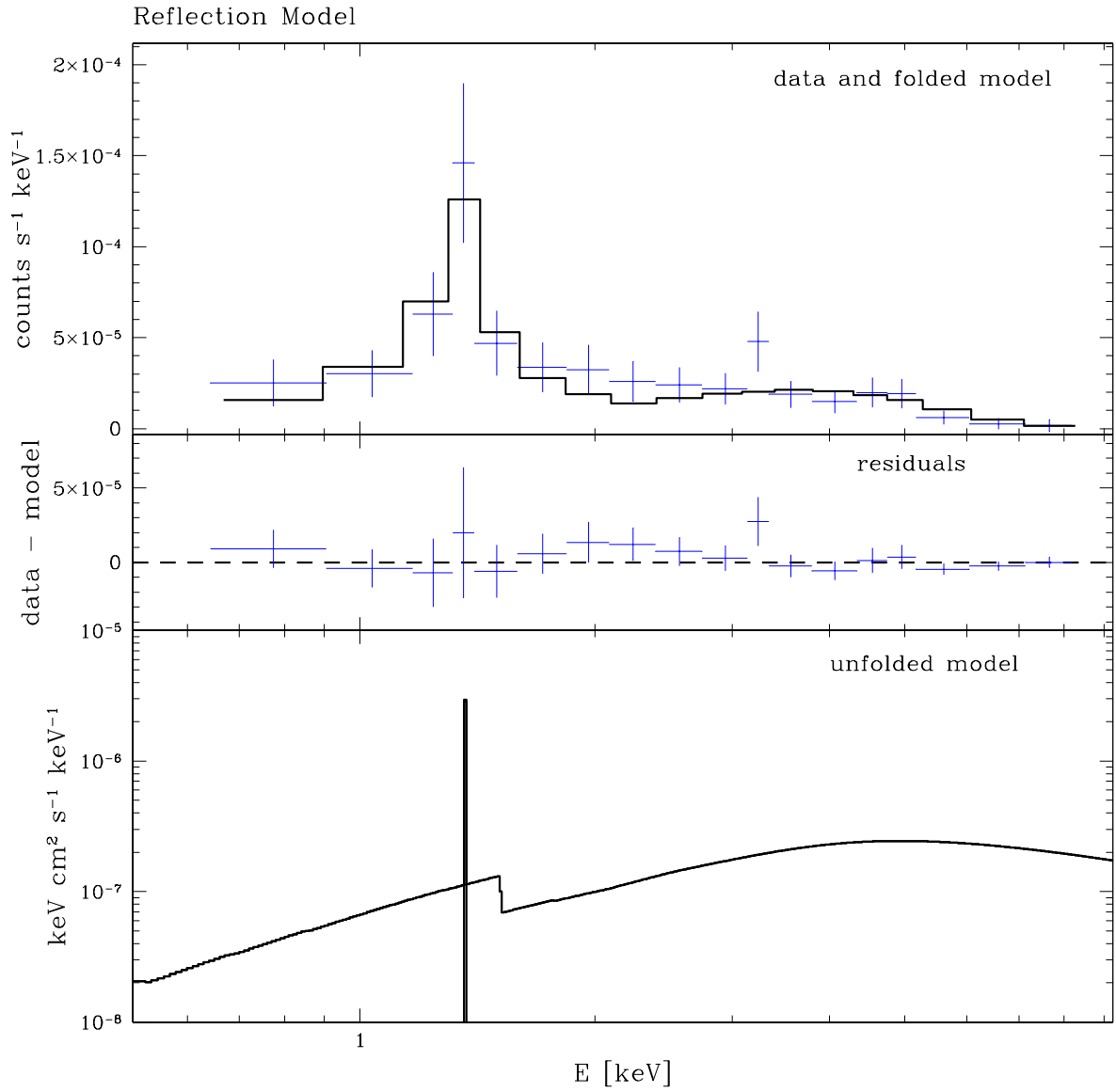}
\caption{Top panels: X-ray Spectrum of CDF--S 202 with Reflection Fit
and residuals.  Bottom panel: the unfolded best--fit model. 
\label{Xreflect}}
\end{figure}

We then considered a pure reflection model using the {\tt pexrav}
model in XSPEC with $\Gamma$ fixed to 1.8 and adding at $\sim 6.4$ keV
rest frame a narrow Gaussian line with $\sigma_{K\alpha}=0$ keV.  The
description of the data is as good as in the transmission model
($\chi^2/dof=10/15$; see Figure~\ref{Xreflect}). With this model the
line is detected at $> 2\sigma$ level according to an $F$-test for one
additional free parameter ($\Delta\chi^2$=4.6 with respect to the same
reflection model without the line). The rest frame line energy and
equivalent width are $E_{\rm K\alpha}=6.43\pm0.34$ keV and $EW_{\rm
K\alpha}=1186^{+1195}_{-922}$ eV, respectively. The line equivalent
width is then in full agreement with the expectations from Compton
thick sources \citep{ghisellini1994}. Assuming a reflection efficiency
of 3\% (see e.g. Krolik, Madau \& Zicky 1994; Maiolino et al. 1998;
and Sect~3.1 in Gilli, Risaliti \& Salvati 1999), the 2--10 keV rest
frame intrinsic luminosity of the source is $1.4\:10^{45}$ erg
s$^{-1}$\footnote{We note that assuming $H_0$=50 km/s/Mpc and
$q_0=0$ gives luminosities higher by a factor of $\sim3$.}.

In principle one could look at the source variability to estimate
which scenario is favored. Indeed, short term variability, on
timescales of a few 10 ksec, indicates that the observed radiation is
transmitted rather than reprocessed by a pc--scale reflecting material.
Unfortunately, analysis of short term variability is not possible
given our photon statistics.  Also, long term variability (the data
were collected in a 14 months period) is poorly constrained.

We also note in passing the blip in the X- ray spectrum at 3.2 keV,
corresponding to a restframe energy $\sim 15$ keV. This is not a
statistically significant line in the spectrum but we point out that
such features have found twice previously in ASCA observations of
other AGN and may be associated with Doppler-bosted $ Fe K\alpha$
\citep{yaqoob1999}.

\section{Discussion and Implications} \label{discussion}

How common are QSO 2s and how can more be detected? This object stood
out in a Hardness Ratio-Redshift plot of the currently cataloged
sources in the {\it Chandra} Deep Field South as the highest redshift
source with the hardest spectrum. One would expect to observe the red
($R-Ks\gtrsim4$) colors of the host galaxy (see e.g. Fig.~7 in Lehmann et
al. 2001). In this case, at $z=3.700$, the $R-Ks$ color of $\sim2.5$
does not reflect the galactic continuum, but rather the chance
super-position of emission lines into the broad--band filters. 
For CDF--S 202, the $R$ band contains both Ly$\alpha$ and CIV, while the
$Ks$ band contains the weaker H$\beta$, H$\delta$ and HeII.
Since the CDF-S 202 optical continuum is not detected, we used the NGC
1068 template redshifted to z=3.7 to verify the effect of line
subtraction to the optical-infrared colors, assuming that the NGC 1068
spectrum is identical to that of CDF-S 202. We found that $R-Ks=3.92$
when lines are subtracted, which is comparable with the Lehmann et
al. findings.

\begin{figure}
\plotone{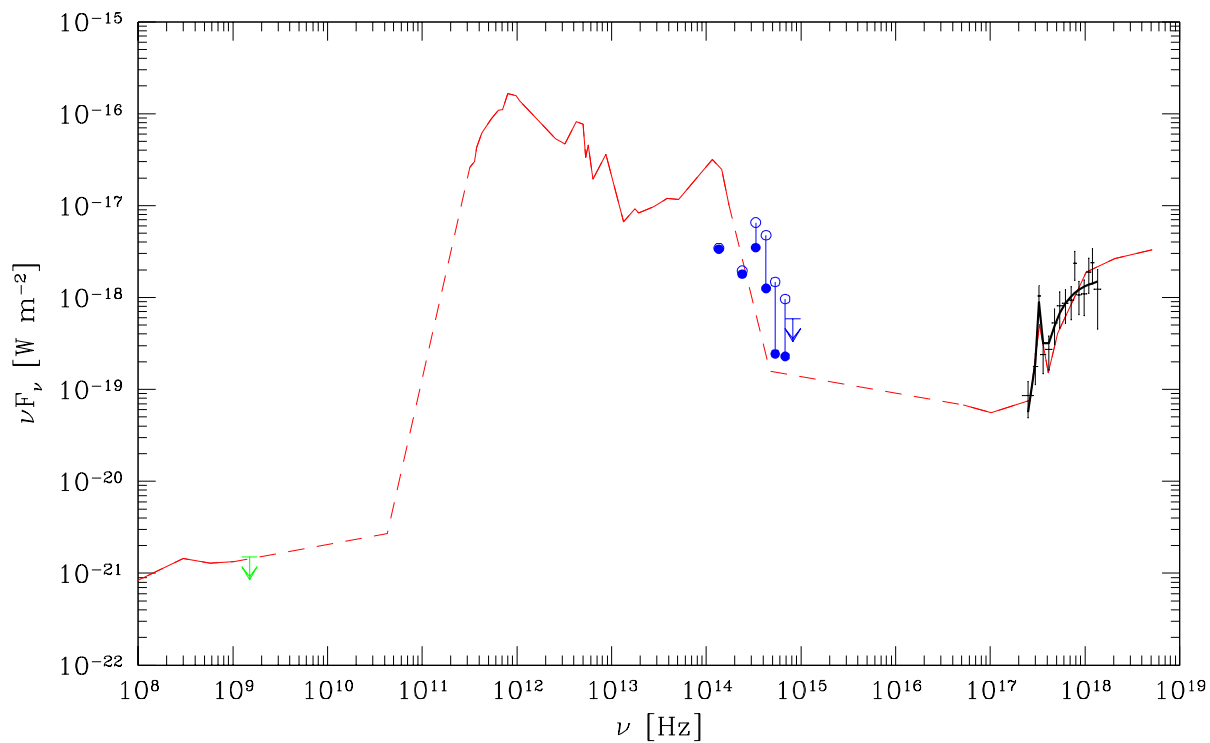}
\caption
{Detections and upper limits for CDF-S 202 plotted over the SED for
the local Type 2 template, NGC 6240 from \citet{hasinger01} (solid
line; dashed line where the template is extrapolated). The CDF-S 202
data have been shifted to the rest-frame. The NGC 6240 SED has been
normalized to the X-ray spectrum of CDF-S 202. Open circles and filled
circles refer to observed magnitudes and to line subtracted
magnitudes, respectively (see the Discussion).
\label{N6240}
}
\end{figure}

Both observed and corrected (i.e. line subtracted) magnitudes are
shown in Fig.~\ref{N6240}, along with the X-ray spectrum and the radio
upper limit ($F_{\nu}<100 \mu Jy$, where $F_{\nu}$ is the flux density
at 20 cm) derived from VLA observations (K.Kellerman, private
communication). To put our data into perspective, they have been
shifted to the rest frame and compared with the SED the nearby galaxy
NGC 6240 (Hasinger 2001), which has no AGN signatures in its optical
spectrum but shows clearly a buried AGN emerging at higher X--ray
energies \cite{vignati}.

The inferred metallicity for this object is $\gtrsim Z_{\odot}$
typical of high redshift AGN but on the high side.  In fact, looking
at our diagnostic diagram, CDF--S 202 has a higher metallicity than
any of the high--z radio galaxies of \citet{Vernet01, hamann93}.
Indeed, it has been found that there is a metallicity evolution such
that for redshift $ z>3$ radio galaxies, the metallicity is $<2\times
Z_{\odot}$ \citep{debreuck00}.  This relatively high metallicity is
consistent with a high star formation rate during these early epochs
as the bulge and central black hole are formed.

Assume now that the $Fe K\alpha$ line arises in the obscuring cloud.
We will now estimate various quantities of the obscuring region using
either standard parameters or, where possible, parameters inferred
from our data. To simplify the calculation we assume that the
obscuring cloud is spherical. From other studies \citep{gilli2001} we
find that the Type 2 to Type 1 ratio is $\sim 10$ at high redshift and
therefore it is a reasonable assumption that the obscuring torus is an
almost totally enclosing cloud.

We define the ionization parameter, U, in the usual manner as

\begin{equation}
U = {L_X \over n r^2}
\end{equation}

in an obscuring material of density, $n$, at distance, $r$ from the
source. For standard values U $= 10^2$ erg cm s$^{-1}$
\citep{matt1996}.

For the inferred absorbing column, N, of the spherical uniform cloud,
we find

\begin{equation}
N = nr   \, .
\end{equation}

For a typical inferred absorbing column we use of $N = 10^{24.5}$
cm$^{-2}$ that is consistent with canonical estimates expected for
Type 2 QSOs and also with our inferred column densities in
Section \ref{x-ray}.

It follows directly that the radius of the cloud is 

\begin{equation}
r = {L_X \over {NU}} \sim 1 {\rm pc}\;(L_X/10^{45} {\rm erg\;s}^{-1})({10^2}
{\rm erg\;cm\;s}^{-1}/U)(10^{24.5}{\rm cm}^{-2}/N) \, \, ,
\end{equation}

and the density of the cloud is

\begin{equation}
n = {{N^2 U} \over{L_X }} .
\end{equation}

The mass estimate of the obscuring cloud, $M_{cloud}$, is

\begin{equation}
M_{cloud} = ({{4 \pi} \over 3}) \mu m_H nr^3 
  =  ({{4 \pi} \over 3}) \mu m_H ({L_x^2 \over{ N U^2}}) 
\end{equation}

where $\mu$ is the mean molecular weight of the obscuring material.
We then estimate the mass in the obscuring region $M_{cloud}$, to be

\begin{equation}
M_{cloud} \sim 1 \times 10^{5} M_{\odot}(L_X/10^{45} {\rm erg\;s}^{-1})
(10^2 {\rm erg\;cm\;s}^{-1}/U)(10^{24.5}{\rm cm}^{-2}/N_H).
\end{equation}

We now assume that the velocity dispersion outside the obscuring
cloud can still be dominated by influence of the black hole. This
will not always be the case but here the numbers are sufficiently
interesting and consistent to analyze the problem further. Then, the
velocity dispersion, $\sigma_{cloud}$, just outside the obscuring region is
given by

\begin{equation}
\sigma_{cloud}^2 = ( {{G M_{BH}} \over r})
\end{equation}

where $M_{BH}$ is the mass of the black hole. It follows directly that 

\begin{equation}
\sigma_{cloud}^2 = ( {{G M_{BH}} \over {L_X}})(NU)
\end{equation}

We further rewrite the black hole mass in terms of the Eddington luminosity
given by 

\begin{equation}
M_{BH} = ({\sigma_T \over {4 \pi G c m_H}}) L_{EDD}
\end{equation}

and then

\begin{equation}
\sigma_{cloud}^2 = ({\sigma_T \over {4 \pi c m_H}})( {{L_{EDD} } \over
 {L_X}})(NU) 
\end{equation} 

Now we can estimate the velocity dispersion, just outside the
obscuring region is

\begin{equation}
\sigma_{cloud} \lesssim 2500  km s^{-1}
[{L/L_{EDD}]}/10^{-2})^{1/2} (U/{10^2} {\rm erg cm
s}^{-1})^{1/2}(N/10^{24.5}{\rm cm}^{-2})^{1/2}.
\end{equation}

We have assumed here that the black hole is generating X--rays with an
efficiency of $1\%$ relative to the Eddington luminosity which we
justify as follows. As discussed by \citet{elvis1994} the ratio of
X--ray luminosity to bolometric luminosity, $L_{bol}$, is 10\%. Here we
assume $L_{bol}/L_{EDD} $ is also 10\%.  The inferred mass of the
black hole is then

\begin{equation}
M_{BH} \sim 8 \times 10^8
M_{\odot}(10^{-2}/[{L_X/L_{EDD}}])(L_X/10^{45} {\rm erg s}^{-1})
\end{equation}

This explains the rather large width of the narrow lines that
originate outside the obscuring region.  That such a dense cloud
closely covers the nucleus in high redshift Type 2 objects is to be
expected since in the galaxy formation process large amounts of gas
are present in the bulge-forming black-hole-growing epoch. The
encapsulation of the nucleus probably increases with redshift.

We now discuss our analysis and analyze the energetics of the
line emission. Following the standard scenario,we assume that:(1) the
ratio of UV to X-ray luminosity is 10 \citep{elvis1994};(2) $10\%$ of
the UV continuum escapes, consistent with the ratio of Type 2/Type 1
$\sim 10$ at high redshift; (3) $10\%$ of the continuum flux is
re-radiated isotropically into lines and (4) the covering factor of
the narrow line region is $10\%$. We find that the line flux should be
$ 10^{-2}L_X$. The actual line flux is $L_{line} \sim 2 \times
10^{43}$erg s$^{-1}$ roughly consistent with the above model. In a
similar calculation we use the observationally derived relation
$L_{Ly\alpha} \sim 10^{-3} L_{bol}$ \citep[T. Heckman, private
communication]{VandenBerk01} where $L_{bol}$ is the bolometric
luminosity of the QSO and $L_{Ly\alpha}$ is
the narrow line luminosity in $Ly\alpha$. We then find that CDF-S 202
has an inferred luminosity from the Ly$\alpha$ line of $L_{bol} = 2
\times 10^{46}$ erg s$^{-1}$. In addition, the X-ray luminosity
inferred from the line strength and the observationally determined
value ($L_x/L_{bol} = 0.1$ \citep{elvis1994} is $L_x = 2 \times
10^{45}$ erg s$^{-1}$. This constructs a consistent picture of CDF-S
as an enshrouded Type2 QSO.

In general for observations to support such an analysis of Type 2 QSOs
physics as given above we would like to study in the future a
significantly lower redshift sample of X-ray selected QSO 2 to attempt
to infer the depth of the potential well via the rotation curve and
the central stellar velocity dispersion using, say, STIS
observations. Imaging observations of such objects in CDF-S with HST
can give important morphological constraints \citep[see
][]{schreier2001} and possibly an estimate of the bulge mass which can
also be used to estimate the black hole mass \citep{magorrian1998,
gebhardt2000}.  In addition, deep {\it Chandra} and XMM observations
of high-redshift Narrow Line Radio Galaxies (NLRG) can give insight
into the physics of Type 2 QSOs. In unified models NLRG are the radio
loud equivalent of Type 2 QSOs. It has been proposed by
\citet{ridgeway2001} that a few percent of the Lyman-break galaxy
population \citep{steidel1999} may be Type 2 QSOs. X-ray studies of a
well-selected subsample of these Lyman-break systems are also
indicated.

It is expected from models that the radiation absorbed in the cold
obscuring torus to be re-radiated in the far infra-red. IRAS
09014+4109 is such an object. Interestingly, it is the only
hyperluminous infrared galaxy to be detected as an obscured AGN
\citep{iwasawa2001}. Deep IR imaging and spectroscopy are essential to
constrain the contribution of QSO 2s to the IR background. Current
thinking is that the AGN contribution may approach that due to star
formation if sufficient Type 2 QSOs contribute. For these powerful
buried QSOs at high redshift the Type2/Type1 ratio is unknown but may
continue to increase with redshift giving a major contribution to the
IR background. In a related issue these Type2 QSOs may be powerful
submillimeter sources.  We already know that a fraction of optically
identified SCUBA sources are AGN \citep{barger1999, smail1999} although
the bulk of the sub-mm population is not likely to be powered by AGN,
as suggested by the non-detections of SCUBA sources with {\it Chandra}
\citep{horn2000}. However, we must await the Atacama Large Millimeter Array
(ALMA) to study them in detail in the CDF-S.

Depending on the assumed cosmology and X-ray transmission or
reflection scenario, the intrinsic 2-10 keV rest frame luminosity of
CDF-S 202 varies from $\sim 3 \times 10^{44}$ to $\sim 3 \times
10^{45}$ erg s$^{-1}$. Even assuming the lowest luminosity value,
CDF-S 202 is a powerful obscured AGN, beyond the knee, or break in the
slope, of the AGN X-ray luminosity function at $L_{*,X} = 10^{43.7}
erg s^{-1}$ \citep{miyaji2000}.Above this break luminosity, $L_{*,X}$
the X-ray selected AGNs are QSOs. Let us now discuss the implications
of CDF-S 202 for the origin of the X--ray background \citep{comastri}.
Based on the population synthesis models described by
\citet{gilli2001}\footnote{Gilli et al. (2001) adopted $H_0$=50
km/s/Mpc, $\Omega_m=1$ and $\Omega_\lambda=0$. Here we quote their
predictions correcting for the different cosmology adopted in this
paper.}, obscured AGN with $L_{2-10}>3 \times 10^{44}$ erg s$^{-1}$
and $N_{\rm H}>10^{22}$ cm$^{-2}$ should make up a significant part
($\sim 30\%$) of the hard X-ray background. About 70\% of them should
be observed at $z=1-3$, while 15\% is expected at higher redshifts,
assuming the AGN space density to decrease above $z\sim3$ as found in
optical and radio surveys (Schmidt, Schneider \& Gunn 1995; Fan et
al. 2001; Shaver et al. 1999). On the same assumption we expect 8
obscured QSO with intrinsic luminosity $L_{2-10}>3\times 10^{44}$ erg
s$^{-1}$, $N_{\rm H}>10^{22}$ cm$^{-2}$ and $z>3$ to be observed in
the 1 Msec CDF-S. We will have to await further analysis of these
objects in CDF-S to understand the contribution of the QSO 2
population relative to the other sources that make up the X-ray
background.

\section{Conclusions and Future Work} \label{conclusions}
 
We conclude that in the {\it Chandra} Deep Field South 1Msec exposure
we have found a QSO 2 at redshift 3.700 (CDF-S 202).  There is no
generally accepted definition of a Type 2 QSO. We suggest that CDF-S
202 defines the class well by optical spectrum, X-ray spectrum and
X-ray luminosity. Its luminosity ($ 10^{45 \pm 0.5}$erg s$^{-1}$ in
the 2--10 keV band), X--ray spectrum, narrow permitted lines ($\sim
1500$ km s$^{-1}$), and undetected continuum (to a limit of $\sim 3
\times 10^{-19}$ erg cm$^{-2}$ s$^{-1}$ \AA$^{-1}$) make it a classic
case. There are several other candidates in the field that have hard
spectra and high redshift. Clearly, in Type 2 QSOs the $Fe K\alpha $
feature can be used in conjunction with deep optical and IR faint
object spectroscopy to determine redshifts. CDF-S 202 has the highest
redshift $Fe K\alpha$ line spectrum yet observed
\citep{reeves2001}. 

We have discussed in the paper the implications for the X-ray
background, the infrared background, the origin of the UV continuum
in Type 2 objects and the relation of the growth of black holes to
bulge and galaxy formation. Currently we aim to both increase the
sample and to observe CDF-S 202 further with: IR spectroscopy and
imaging, HST, millimeter facilities and XMM.

\acknowledgements

C.N. thanks Tim Heckman, Julian Krolik, David Neufeld, David
Strickland and Tahir Yaqoob for stimulating discussions. G. Hasinger
acknowledges support under DLR grant 50 OR 9908 O.  R. Giacconi and
C. Norman gratefully acknowledge support under NASA grant NAG-8-1527
and NAG-8-1133.

\clearpage

\newpage

\clearpage

\begin{center}
\begin{deluxetable}{lccccccc}
\tablecolumns{8} \tablecaption{CDF-S 202 Source Properties}
\tabletypesize{\scriptsize} \tablewidth{0pt} \tablehead{\colhead{} &
\colhead{} & \colhead{} & \colhead{} & \colhead{} & \colhead{} &
\colhead{} & \colhead{}} \startdata 
\sidehead{{\bf X-ray Data}} 
Model& Transmission& Reflection& &&&&\\ 
$\Gamma$(fixed)& 1.8& 1.8& &&&&\\
$N_H$ (10$^{23}$ cm$^{-2}$)& $7.9_{-2.3}^{+3.5}$ & $\geq10^{2}$&&&&&\\
$\textrm{EW}^a_{\textrm{\tiny Fe K}\alpha}$ (rest-frame eV) &
$823_{-823}^{+1694}$& $1186_{-922}^{+1195}$ &&&&&\\ 
$F^b_{0.5-2}$ (10$^{-16}$ erg s$^{-1}$cm$^{-2}$) & $2.21\pm0.33$& $2.25\pm0.33$&&&&&\\ 
$F^b_{2-10}$ (10$^{-16}$ erg s$^{-1}$cm$^{-2}$) & $20.6\pm3.0$& $25.8\pm3.5$&&&&&\\ 
\hline
\sidehead{{\bf Optical Imaging}} Band & $U$ & $B$ & $V$ & $R$ & $I$ &
$J$ & $Ks$\\ Vega magnitudes & $>26$ & $26.2\pm0.3$ & $25.27$ &
$23.53$ & $22.65\pm0.1$ & $23.28\pm0.15$ & $20.97\pm0.11$\\ \hline
\sidehead{{\bf Optical Spectroscopy (rest frame) }} 
& Ly$\alpha$& NV& CIV& HeII & && \\ 
F ($10^{-18}$ erg s$^{-1}$ cm$^{-2}$)$^c$ & 16.4 & 5.9 & 9.9 & 2.8 & &&\\ 
FWHM (\AA) & $<$5 & $<$7 & $<$9 & $<$4 & &&\\ 
Velocity Width (km s$^{-1}$) & $<$1130 & $<$1680 & $<$1680 & $<$680 & && \\ \hline
\sidehead{{\bf IRAS Imaging}} Band & $S_{12\mu}$ & $S_{25\mu}$ &
$S_{60\mu}$ & $S_{100\mu}$&&&\\ $F_{\nu}$ (Jy) & $<0.2$ & $<0.2$ &
$<0.2$ & $<1.0$ &&&\\ \hline \sidehead{{\bf VLA 20cm Imaging}}
$F_{\nu}$ ($\mu$Jy) & $<100$ &&&&&&\\ \enddata

$^a$ Assuming 6.4 keV and 0 keV for the rest frame line energy and width, respectively.\\
$^b$ Observed flux.\\
$^c$ Errors in emission-line fluxes are estimated to be 5\%.
\end{deluxetable}
\end{center}
\end{document}